\documentclass[twocolumn]{aastex62}
\usepackage{verbatim}

\newcommand {\hi}{\ifmmode \left{\rm H}\,{\textsc i}\right \else H\,{\sc i}\fi}
\newcommand{\kms}{$\rm km~s^{-1}$}

\newcommand {\oii}{\ifmmode \left[{\rm O}\,{\textsc ii}\right] \else [O\,{\sc ii}]\fi}
\newcommand {\OIII}{\ifmmode \left[{\rm O}\,{\textsc iii}\right] \else [O\,{\sc iii}]\fi}
\newcommand {\SII}{\ifmmode \left[{\rm S}\,{\textsc ii}\right] \else [S\,{\sc ii}]\fi}

\newcommand {\heii}{\ifmmode {\rm He}\, {\sc II\lambda4686}\ \else He\,{\sc II$\lambda$4686}\fi}
\newcommand {\civ}{\ifmmode {\rm C}\, {\sc IV}\ \else C\,{\sc IV}\fi}
\newcommand {\hb}{\ifmmode {\rm H}\beta \else H$\beta$\fi}
\newcommand {\ha}{\ifmmode {\rm H}\alpha \else H$\alpha$\fi}

\newcommand {\hei}{\ifmmode {\rm He} {textsc{i}} \else He\,{\sc i}\fi}
\newcommand {\mgii}{\ifmmode {\rm Mg}{\textsc{ii}} \else Mg\,{\sc ii}\fi}
\newcommand {\mgiii}{\ifmmode {\rm Mg}{\textsc{iii}} \else Mg\,{\sc iii}\fi}

\newcommand{\erg}{${\rm erg \ s^{-1}}$ }

\newcommand{\ergcms}{\ifmmode {\rm ergs\,cm}^{-2}\,{\rm s}^{-1} \else ergs\,cm$^{-2}$\,s$^{-1}$\fi}
\newcommand{\ergcmsA}{\ifmmode{\rm ergs}\, {\rm cm}^{-2}\,{\rm s}^{-1}\,{\rm\AA}^{-1} \else ergs\, cm$^{-2}$\, s$^{-1}$\, \AA$^{-1}$\fi}
\newcommand{\cmz}{${\rm cm^{-2} }$ }
\newcommand{\cmt}{${\rm cm^{-3} }$ }
\usepackage{CJK}
\usepackage{xcolor,color}
\usepackage{hyperref}
\hypersetup{
   colorlinks,
   linkcolor={blue!88!black!80},
   citecolor={blue!88!black!80},
   urlcolor={blue!88!black!80}
}
\shorttitle{LOC model }
\shortauthors{Guo et al.}
 
\begin{document}
\begin{CJK}{UTF8}{gbsn}

\title{Understanding Broad \mgii\ Variability in Quasars with Photoionization: Implications for Reverberation Mapping and Changing-Look Quasars}

\author[0000-0001-8416-7059]{Hengxiao Guo (郭恒潇)}

\affiliation{Department of Astronomy, University of Illinois at Urbana-Champaign, Urbana, IL 61801, USA}
\affiliation{National Center for Supercomputing Applications, University of Illinois at Urbana-Champaign, Urbana, IL 61801, USA}

\author[0000-0003-1659-7035]{Yue Shen}
\altaffiliation{Alfred P. Sloan Research Fellow}
\affiliation{Department of Astronomy, University of Illinois at Urbana-Champaign, Urbana, IL 61801, USA}
\affiliation{National Center for Supercomputing Applications, University of Illinois at Urbana-Champaign, Urbana, IL 61801, USA}

\author{Zhicheng He}
\affiliation{CAS Key Laboratory for Researches in Galaxies and Cosmology, University of Sciences and Technology of China, Hefei, Anhui 230026, China}
\affiliation{School of Astronomy and Space Science, University of Science and Technology of China, Hefei 230026, China}

\author[0000-0002-1517-6792]{Tinggui Wang}
\affiliation{CAS Key Laboratory for Researches in Galaxies and Cosmology, University of Sciences and Technology of China, Hefei, Anhui 230026, China}
\affiliation{School of Astronomy and Space Science, University of Science and Technology of China, Hefei 230026, China}

\author[0000-0003-0049-5210]{Xin Liu}
\affiliation{Department of Astronomy, University of Illinois at Urbana-Champaign, Urbana, IL 61801, USA}
\affiliation{National Center for Supercomputing Applications, University of Illinois at Urbana-Champaign, Urbana, IL 61801, USA}

\author{Shu Wang}
\affiliation{Kavli Institute for Astronomy and Astrophysics, Peking University, Beijing 100871, China}
\affiliation{Department of Astronomy, School of Physics, Peking University, Beijing 100871, China}

\author[0000-0002-0771-2153]{Mouyuan Sun}
\affiliation{CAS Key Laboratory for Researches in Galaxies and Cosmology, University of Sciences and Technology of China, Hefei, Anhui 230026, China}
\affiliation{School of Astronomy and Space Science, University of Science and Technology of China, Hefei 230026, China}

\author[0000-0002-6893-3742]{Qian Yang}
\affiliation{Department of Astronomy, University of Illinois at Urbana-Champaign, Urbana, IL 61801, USA}
\affiliation{National Center for Supercomputing Applications, University of Illinois at Urbana-Champaign, Urbana, IL 61801, USA}

\author{Minzhi Kong}
\affiliation{Department of Physics, Hebei Normal University, No. 20 East of South 2nd Ring Road, Shijiazhuang 050024, China}
\affiliation{Department of Astronomy, University of Illinois at Urbana-Champaign, Urbana, IL 61801, USA}

\author[0000-0001-6938-8670]{Zhenfeng Sheng}
\affiliation{CAS Key Laboratory for Researches in Galaxies and Cosmology, University of Sciences and Technology of China, Hefei, Anhui 230026, China}
\affiliation{School of Astronomy and Space Science, University of Science and Technology of China, Hefei 230026, China}

\email{hengxiao@illinois.edu (HG), shenyue@illinois.edu (YS)}

\begin{abstract}
The broad \mgii\ line in quasars has distinct variability properties compared with broad Balmer lines: it is less variable, and usually does not display a ``breathing'' mode, the increase in the average cloud distance when luminosity increases. We demonstrate that these variability properties of \mgii\ can be reasonably well explained by simple Locally Optimally Emitting Cloud (LOC) photoionization models, confirming earlier photoionization results. In the fiducial LOC model, the \mgii-emitting gas is on average more distant from the ionizing source than the \ha/\hb\ gas, and responds with a lower amplitude to continuum variations. If the broad-line region (BLR) is truncated at a physical radius of $\sim 0.3$ pc (for a $10^{8.5}M_{\odot}$ BH accreting at Eddington ratio of 0.1), most of the \mgii\ flux will always be emitted near this outer boundary and hence will not display breathing. These results indicate that reverberation mapping results on broad \mgii, while generally more difficult to obtain due to the lower line responsivity, can still be used to infer the \mgii\ BLR size and hence black hole mass. But it is possible that \mgii\ does not have a well defined intrinsic BLR size-luminosity relation for individual quasars, even though a global one for the general population may still exist. The dramatic changes in broad \ha/\hb\ emission in the observationally-rare changing-look quasars are fully consistent with photoionization responses to extreme continuum variability, and the LOC model provides natural explanations for the persistence of broad \mgii\ in changing-look quasars defined on \ha/\hb, and the rare population of broad \mgii\ emitters in the spectra of massive inactive galaxies. 
\end{abstract}



\section{Introduction}\label{sec:intro}

The ubiquitous aperiodic variability of quasars can be utilized to probe different spatial scales, for example, to measure the size of the broad-line region (BLR) and hence to estimate the black hole (BH) mass. The origin of quasar continuum variability could be changes in the accretion rate \citep[e.g.,][]{Li08}, or complex disk instabilities \citep[e.g.,][]{Lyubarskii97,Dexter11,Cai16}. The delayed responses of more extended emitting regions (such as the BLR) to the continuum variations measure the characteristic distance of these emitting regions to the central BH ($R= c\tau$, where $\tau$ is the time lag between continuum and broad line variability and $c$ is the speed of light), a technique known as reverberation mapping \citep[RM,][]{Blandford82,Peterson14}. Combined with the broad line width $\Delta V$ (a proxy for the virial velocity in the BLR) measured from spectroscopy, one can estimate the BH mass as 
\begin{equation}\label{eq:gravity}
\rm M_{\rm BH} = \frac{\it f \Delta V^2 R}{\it G}\ ,
\end{equation}
where $f$ is a scale factor of order unity that accounts for the BLR orientation, kinematics, structure and other unknown factors, and $G$ is the gravitational constant. So far there have been more than 60 low-redshift ($z < 0.3$) AGNs with successful RM lag measurements \citep[e.g.,][]{Kaspi00,Peterson04,Barth15,Du16} (see a recent compilation of the RM BH mass database from \citet{Bentz15}), mostly for the broad \hb\ line. These measurements have revealed a scaling relation between the size of the BLR and the continuum luminosity (i.e., $R\propto L^{0.5}$) \citep[e.g.,][]{Kaspi00,Bentz06} for \hb, which is naively expected from photoionization, i.e.,  $R\propto L_{\rm ion}^{0.5}$, where $L_{\rm ion}$ is the hydrogen ionizing luminosity in photoionzation models \citep[e.g.,][]{Korista00}. This $R-L$ relation underlies the BH mass estimation with single epoch spectroscopy, using luminosity as a proxy for the BLR size \citep[e.g.,][]{Shen13}.

At $1\lesssim z\lesssim 2$, the strong Balmer lines \ha\ and \hb\ shift out of the optical band, and \mgii\ becomes the major broad line of interest for RM with optical spectroscopy. Compared with the Balmer lines, results on \mgii\ RM are scarce and more ambiguous. Only in a handful cases has a \mgii\ lag been robustly detected \citep{Clavel91,Reichert94,Metzroth06,Shen16,Czerny19}, with many attempts failed to result in a detection \citep{Trevese07,Woo08, Hryniewicz14,Cackett15}. Part of the difficulty of \mgii\ lag detection is the apparently lower variability amplitude of \mgii\ compared to the Balmer lines in the same objects \citep[e.g.,][]{Sun15}. 

Extensive earlier theoretical works on photoionization have suggested that \mgii\ not only has a weaker response to continuum variations resulting in weaker line variability, but also a larger average formation radius, compared to Balmer lines \citep{Goad93,O'Brien95,Korista00,Goad12}. Based on the IUE (International Ultraviolet Explorer) monitoring of the Seyfert 1 galaxy NGC 4151, \cite{Metzroth06} reported a reliable \mgii\ time lag ($\tau$ $\sim$ 5-7 days) but weak line variability -- the fractional \mgii\ line variability is less than $30\%$ of that for the continuum around 1355 \AA. Similarly, \citet{Clavel91} detected a marginal \mgii\ lag in NGC 5548 with weak line variability (the fractional \mgii\ variability is only $\sim 30\%$ with respect to continuum). \cite{Cackett15} also attempted \mgii\ RM in NGC 5548 with Swift spectra sampled every two days in 2013. However, there was no significant \mgii\ lag detected given the weak \mgii\ variability. Other studies of individual objects also confirmed the general weak variability of \mgii, e.g., NGC 3516 \citep{Goad99a,Goad99b}, PG 1634+706 and PG 1247+268 \citep{Trevese07}, and CTS C30.10 \citep{Modzelewska14}. In addition, population studies of quasar variability in the SDSS Stripe 82 and SDSS-RM \citep{Shen15} also found that \mgii\ variability is generally weaker than those from Balmer lines \citep{Kokubo14,Sun15}. Studies of the broad line responses to large-amplitude continuum variations (more than 1 magnitude) over multi-year timescales also confirmed that the \mgii\ response to continuum is much weaker than that for the broad Balmer lines \citep[][]{Yang19}. Meanwhile, the predicted relatively larger \mgii\ formation radius than the Balmer lines is also consistent with limited RM results where both the \mgii\ and Balmer line lags are available \citep[e.g.,][]{Clavel91,Peterson99,Shen16,Grier17}.

On the other hand, intense RM monitoring of the broad Balmer lines has revealed a ``breathing mode'' of the line \citep{Gilbert03,Korista04b,Cackett06,Denney09,Park12,Barth15,Runco16}: as the central luminosity increases more distant clouds can be photoionized to produce broad line emission, and the broad line width (inversely related to the emissivity-weighted radius) decreases, and vice versa.
Assuming the BLR is virialized, we expect that the line width $\Delta V\propto$ $L^{-0.25}$ assuming $R$ $\propto$ $L^{0.5}$ as for the \hb\ BLR.

The ``breathing'' mode for broad \mgii\ has not been well studied. Recently, \cite{Yang19} studied 16 extreme variability quasars with spectroscopy covering \mgii, and found that the line width of \mgii\ does not vary accordingly as continuum varies by more than a factor of few for most objects (See their Fig. 4), in contrast to the well-known ``breathing" model for \hb. Similar results are reported in \citet{Shen13} for a large sample of quasars with two-epoch spectroscopy to probe the continuum and broad line variability. This is somewhat surprising, given that \mgii\ is also a low-ionization line like the Balmer lines, and that the average \mgii\ width correlates well with that of \hb\ for the population of quasars \citep{Shen08,Shen11}. In rare cases reported, however, \mgii\ may also display breathing, albeit to a lesser degree compared to \hb\ \citep[e.g.,][]{Dexter19a}.

Given the importance of broad \mgii\ for RM at intermediate redshifts and for understanding quasar BLR in general, it is important for us to understand \mgii\ variability. Unlike the recombination lines (e.g., \hb\ and \ha), \mgii\ is mostly collisionally excited. To fully understand the different variability properties of broad \mgii, detailed photoionization calculations are required. In this work we construct photoionization models using {\tt CLOUDY} \citep[][version 17.01]{Ferland17} to understand \mgii\ variability and compare to the Balmer lines. In \S \ref{sec:mechanism}, we discuss the excitation and emission mechanisms of different lines. 
In \S \ref{sec:LOC} we confirm the low intrinsic response and large formation radius for \mgii\ w.r.t. Balmer lines in the LOC picture \citep[e.g.,][]{Goad93,O'Brien95,Korista00,Goad12}. In \S \ref{sec:imp}, we discuss the implications of \mgii\ variability on RM, and present a sequence of ``changing-look'' in Balmer lines and \mgii\ as the continuum luminosity undergoes significant variations. We summarize our findings in \S \ref{sec:con}.      

\section{Emission Mechanisms for \ha, \hb\ and \mgii}\label{sec:mechanism}

In order to investigate the variability behaviors of the three different broad lines, we first need to understand their excitation and emission mechanism and response to the central ionizing source. \ha\ and \hb\ are recombination lines in the BLR with large column density ($\sim$ $10^{23}$ \cmz) and high volume density ($\sim$ $10^{10}$ \cmt) clouds. The ionization potential of \hi\ is 13.6 eV, and the ionized \hi\ will subsequently recombine and emit Balmer lines (e.g., \hb\ and \ha) in ionization equilibrium. The recombination timescale is given by 
\begin{equation}\label{eq:rec}
\tau_{\rm rec} = (n_{\rm e}\alpha)^{-1} \sim 0.1( 10^{10}\, \rm cm^{-3} /\it n_{\rm e})\ \rm hr,
\end{equation}
which is related to the recombination coefficient $\alpha$ and the electron density $n_{\rm e}$. Assuming $n_{\rm e}$ = $10^{10}$ \cmt, it yields a recombination timescale of a few minutes.

\begin{figure}
\centering
\includegraphics[width=8.cm]{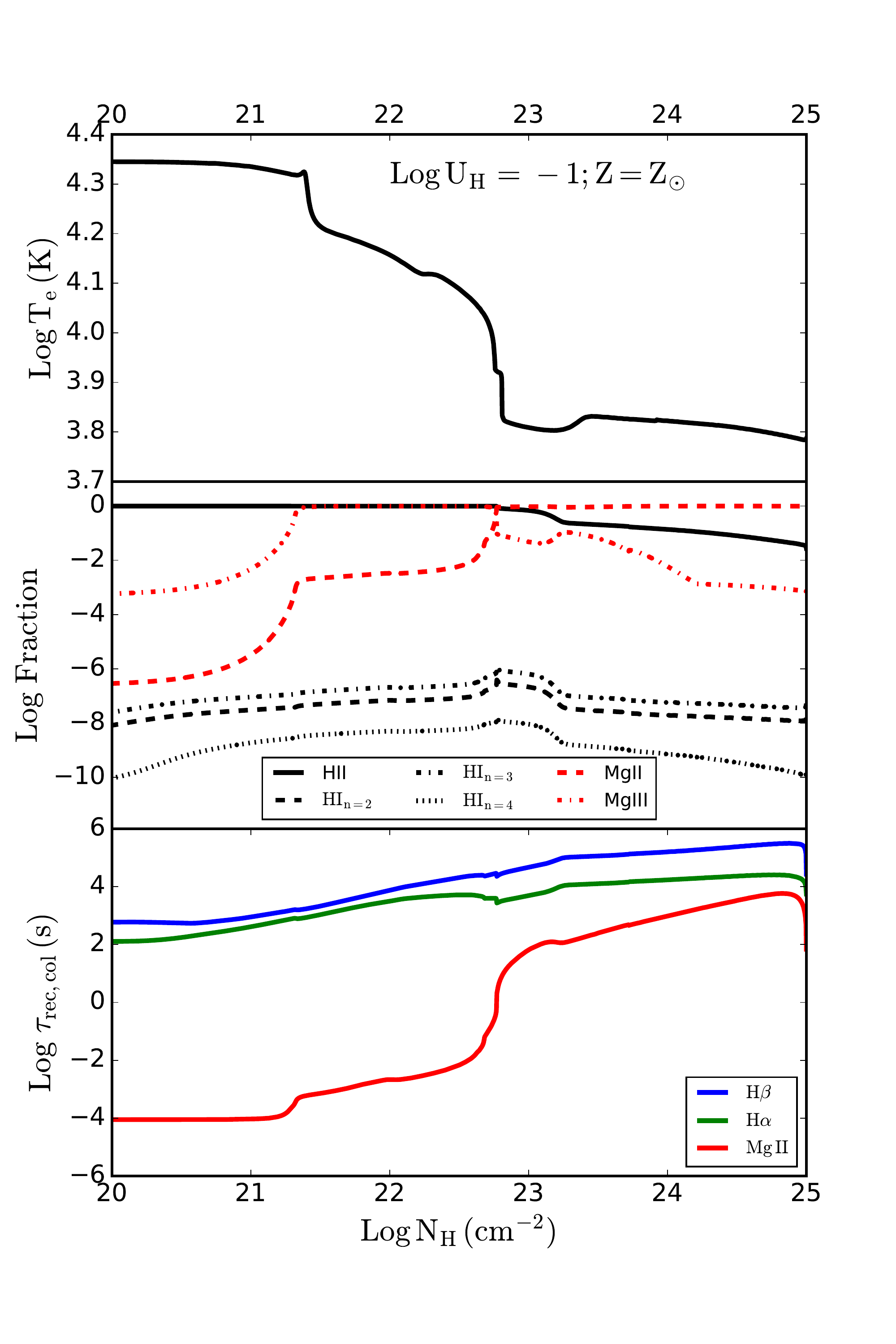}
\caption{Electron temperature, atom/ion number density fractions and excitation timescales as a function of column density. We adopt a simple one-cloud model with typical BLR parameters, i.e., ionization parameter $\log U_{\rm H}=-1$ and chemical abundance $Z= Z_{\odot}$. We perform radiative transfer calculations up to $N_{\rm H}$ = $10^{25}$ \cmz. In the top panel, the electron temperature decreases with increasing hydrogen column density. The middle panel shows the ionization states of hydrogen and magnesium as a function of total column density. \mgii\ has its significant production around $N_{\rm H}$ = $10^{23}$ \cmz since its abundance (emitting efficiency) increases (decreases) with the hydrogen column density.}  The bottom panel shows the recombination timescales of \hb\ and \ha\ and collisional timescale of \mgii, which are consistent with the estimation from Eqns. (\ref{eq:rec}) \& (\ref{eq:col}). 

\label{fig:depth}
\end{figure}

Unlike Balmer lines, \mgii\ is dominated by collisional excitation given its low excitation energy (4.4 eV). The ionization energy of \mgii\ to \mgiii\ is $>$ 15 eV (comparable to the 13.6 eV ionization energy of \hi), which indicates a similar emissivity-weighted radius as the Balmer lines. Its recombination rate is also similar to that of the Balmer lines, whereas the magnesium abundance ($Z$ = log $N_{\rm Mg_{\sc\,II}}/N_{\rm H}$ + 12 = 7.58, where $N$ is the column density of different elements) is four orders of magnitude lower than hydrogen ($Z$ = $Z_{\odot}$). Therefore the contribution from recombination is negligible for \mgii\ emission.  The collisional timescale of \mgii\ is given by 
\begin{equation}\label{eq:col}
\tau_{\rm col} =  \bigg({\frac{8.63 \times 10^{-6} n_{\rm e} \gamma_{\rm mn}  {\rm exp}(-\chi_{\rm mn}/kT) }{\omega_{m}T_{\rm e}^{0.5}}  \bigg)^{-1}} \,\rm s,
\end{equation}
where $\gamma_{\rm mn}=16.9$ is the collisional strength for $3p\, ^{2}P^{0} - 3s\, ^{2}S^{0}$ transitions,  $k$ is the Boltzmann constant, the statistical weight $\omega_m$ is 6 for $3p\, ^{2}P^{0}$, and $\chi_{\rm mn} = E_{\rm m} - E_{\rm n}$ = 4.4 eV for \mgii\ resonance transition \citep{Osterbrock89}. Assuming gas electron temperature $T_{\rm e}=10^{4}$ K and $n_{\rm e}=10^{10}$ \cmt, we obtain the approximate collisional timescale of $\sim$ 0.01s.  

Fig. \ref{fig:depth} presents various cloud parameters as a function of column density in the simple one-cloud model. We assume a slab cloud in an open geometry, with ionization parameter $\log U_{\rm H} = -1$, a column density of $N_{\rm H}=10^{25}$ \cmz, and solar abundance $Z = Z_{\rm \odot}$, is irradiated on one side by a typical radio-quiet AGN SED with the big blue bump \citep{Mathews87}. As column density increases, the electron temperature drops (top panel), and \mgii\ abundance starts to exceed that of \mgiii\ above $N_{\rm H}\sim 10^{22.7}$ \cmz, i.e., \mgii\ is not over-ionized at sufficiently high column densities (middle panel). However, the \mgii\ emitting efficiency will decrease at very high hydrogen column density (e.g., $10^{25}$ $\rm cm^{-2}$ corresponds to Compton thick clouds), resulting in that \mgii\ emission reaches peak production around $N_{\rm H}\sim 10^{23}\, {\rm cm^{-2}}$. The hydrogen abundances at different energy levels are relatively constant. In the bottom panel, we show that the timescales ($\tau_{\rm rec}$ of recombination and $\tau_{\rm col}$ of collisional excitation) are much smaller than 1 day, and hence negligible compared with the time delay in BLR response or the interval of multi-epoch observations.   

In the next section, we use photoionization results calculated for individual clouds to construct more realistic BLR models that cover a distribution in cloud properties with a spherically symmetric geometry.
 
\begin{figure*}
\centering
\hspace*{-1cm} 
\includegraphics[width=20.cm]{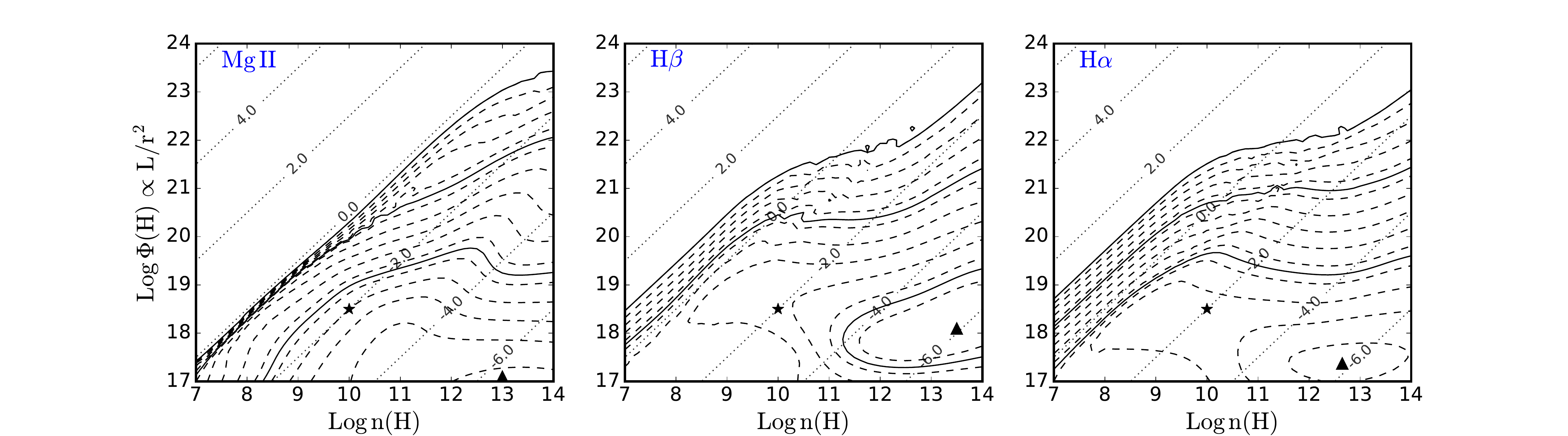}
\caption{Contours of $\rm LogEW$ for three emission lines are shown as a function of the hydrogen density $n(H)$ [\cmt] and surface flux of hydrogen ionizing photons $\Phi \propto L/r^{2} $. We assume the typical hydrogen column density $n(H)$ = $10^{23}$ \cmz, solar abundance ($Z_{\odot}$). The line EW is referenced to the incident continuum at 1215\AA\ for direct comparisons with earlier work, which is directly proportional to the continuum reprocessing efficiency. The smallest decade contour (outermost) corresponds to 1 \AA, and each solid line is 1 decade, and dashed lines represent 0.125 decade steps. The dotted diagonal lines are photoionization parameters decreasing from the upper left ($\log U = 6$) to the lower right ($\log U= -6$). The triangles and stars are the EW peaks and the BLR parameters used in \citet{Davidson79}, respectively.}
\label{fig:LOC}
\end{figure*}

\section{LOC Models}\label{sec:LOC}

A physically-motivated photoionization model for the BLR is the Locally Optimally Emitting Cloud (LOC) model, which consists of clouds with different gas densities and distances from the central continuum source with an axisymmetric distribution \citep{Baldwin95}. In this model, the line emission we observe originates from the combination of all clouds but is dominated by those with the highest efficiency of reprocessing the incident ionizing continuum, i.e., those clouds with the optimal distance from the central source and gas density. This natural selection effect is due to a combination of ionization potential, collisional de-excitation of the upper levels, and thermalization at large optical depths.

In the follow sections, we will consider a typical quasar at $z = 0.5$ with $M_{\rm BH}$ = $10^{8.5}$ $M_{\odot}$ and $L_{\rm 3000\AA}$ = $10^{44-45}$ \erg\ (or Q(H) $\simeq$ $10^{54.5-55.5}\ (\rm s^{-1}))$\footnote{$\rm log\, \it Q(\rm H) = log \, \it L_{\rm bol} + A$, where A is about 9.8 \citep{Arav13} and log$L_{\rm bol}$ = $\rm logBC_{\rm 3000\AA}$ + log$L_{\rm 3000\AA}$, where $\rm logBC_{\rm 3000\AA}$ = 0.71 is the bolometric correction at 3000\AA\ using the average quasar spectral energy distribution from \citet{Richards06}.}, corresponding to Eddington ratios $ L_{\rm bol}/L_{\rm Edd}\sim 0.01\ \rm to\ 0.1$. The high Eddington ratio ($L_{\rm bol}/L_{\rm Edd}\sim 0.1$) represents the typical Eddington ratio observed in quasars \citep[e.g.,][]{Shen11}, and the low Eddington ratio ($L_{\rm bol}/L_{\rm Edd}\sim 0.01$) represents the state where the quasar has significantly dimmed in accretion luminosity. We still use the same radio-quiet AGN SED in \citet{Mathews87} for the incident SED. 


Consider the best studied AGN NGC 5548 \citep[e.g.,][]{Korista00}, which has an average Q(H) = $10^{54.13}\,{\rm s^{-1}}$ and an outer BLR boundary of about 140 lt-day, determined by dust sublimation of the inner edge of the torus with a surface ionizing flux $\rm log\,\Phi(H) \simeq 17.9\ (cm^{-2}s^{-1})$\citep{Nenkova08,Landt19}. Scaling to the average luminosity of $10^{44.5}$ \erg\ at 3000\AA\ for our quasar, we determine an outer BLR boundary of $R_{\rm out} = 10^{18}\, \rm cm$ (since $\rm Q(H) = 4\pi R^2 \Phi(H)$, see Eqn. \ref{eq:u}). For completeness, the other parameters used in our fiducial LOC model are: the inner BLR boundary $R_{\rm in} = 10^{16.5}$ cm, the hydrogen column density $N_{\rm H} =10^{23}$ $\rm cm^{-2}$, a radio-quiet incident AGN SED, solar abundance $Z = Z_{\rm \odot}$, a total cloud covering factor ${\rm CF} = 50\%$ and a cloud distribution as a function of radius and volume density $f(r) \propto r^{\Gamma}$ with $\Gamma$ = $-1.1$ and $g(n) \propto n^{\beta}$ with $\beta = -1$. We justifiy these parameters in detail in the following sections.     

\subsection{Responses of \ha, \hb, and \mgii\ to continuum variations}\label{sec:weakmgii}

Following previous works in a spherically symmetric geometry \citep{Korista00,Korista04b}, we assume that the density of the clouds covers a broad range of 7 $\leq$ $\rm log\, n(H)$ [\cmt] $\leq$ 14 with 0.125 dex spacing. The upper limit is subject to model uncertainties \citep{Ferland13}, and the lower limit is determined by the absence of forbidden lines in the BLR. An ionizing flux range of 17 $\leq$ $\rm log\, \Phi(H)$ [\cmz $\rm s^{-1}$] $\leq$ 24 with 0.125 dex spacing is chosen. The lower limit is related to the sublimation temperature ($\sim$ 1500 K) of dust grains, which corresponds to a hydrogen ionizing flux of $\Phi(H)$ $\sim$ $10^{17-18}$ $\rm cm^{-2}\,s^{-1}$ \citep{Nenkova08}. The upper limit is unimportant since the emissivity for most lines is low at large $\Phi(H)$. The ionization parameter is defined as the ratio of hydrogen-ionizing photon $Q(H)$ to total hydrogen density $n(H)$: 
\begin{equation}\label{eq:u}
U_{\rm H} \equiv \frac{Q(\rm H)}{4\pi R^2 n(\rm H) c} \equiv \frac{\Phi(\rm H)}{n(\rm H) c}\ ,\\
\end{equation}
where $R$ is the distance between the central ionizing source and the illuminated surface of the cloud, $c$ is the speed of light, and $\Phi(H)$ is the flux of ionizing photons. Constant $U_{\rm H}$ values are thus diagonal lines in the density--flux plane of Fig.\ \ref{fig:LOC} (in log-log scale) ranging from 6 (upper left) to $-6$ (lower right). For each cloud (represented by a point in the density-flux plane), we assume a hydrogen column density $N_{\rm H}\, =\, 10^{23}$ \cmz, abundance $Z = Z_{\rm \odot}$ and the same radio-quiet AGN SED \citep{Mathews87} to perform the photoionization calculations using {\tt CLOUDY} \citep[version 17.01,][]{Ferland17}. In addition, we assume an overall covering factor of ${\rm CF}= 50\%$, as adopted in \cite{Korista04a}. Note that the incident continuum does not include the reprocessed emission from other clouds, nor do we consider the effects of cloud-cloud shadowing or continuum/line beaming. 

We obtain the photoionization results on a grid of the density-flux plane, which include 3249 model calculations in Fig. \ref{fig:LOC}, where the contour represents the line equivalent width (EW, w.r.t. the incident continuum at 1215\AA, for direct comparisons with earlier photoionization work) starting from EW = 1 \AA\ in the upper left to $\sim$1000 \AA\ in the lower right. The EW is directly proportional to the continuum reprocessing efficiency for each line (see the figure 4 in \cite{Korista04a}). For constant flux, the vertical y-axis in Fig. \ref{fig:LOC} is equivalent to the radius axis. For all three lines, the most efficiently emitting regions are located in the lower right half, whereas the top left half are regions of Comptonization (the clouds are transparent to the incident continuum). The EW peaks are marked by black triangles while the stars correspond to the old standard BLR parameters \citep{Davidson79}. The peak locations are determined by atomic physics and radiative transfer within the large range of line-emitting clouds. We can see that qualitatively the most-efficient \mgii-emitting clouds is slightly more distant than those for the Balmer lines {given $\Phi(H) \propto L_{\rm ion}/r^2$ (hereafter $L/r^2$)}.       

Next, we compute the overall line luminosity by summing over all grid points with proper weights determined from assumed distribution functions. Following \cite{Baldwin95}, the observed emission line luminosity is the integration given by
\begin{equation}\label{eq:Lline}
L_{\rm line} \propto \int\int^{R_{\rm out}}_{R_{\rm in}} r^2 F(r) f(r) g(n) dn dr \ ,
\end{equation}
where $F(r)$ is the emission line flux of a single cloud at radius $r$,  and $f(r)$  and $g(n)$ are the assumed cloud covering fractions as functions of distance from the center and gas density, respectively. 

\begin{figure*}
\centering
\includegraphics[width=18.cm]{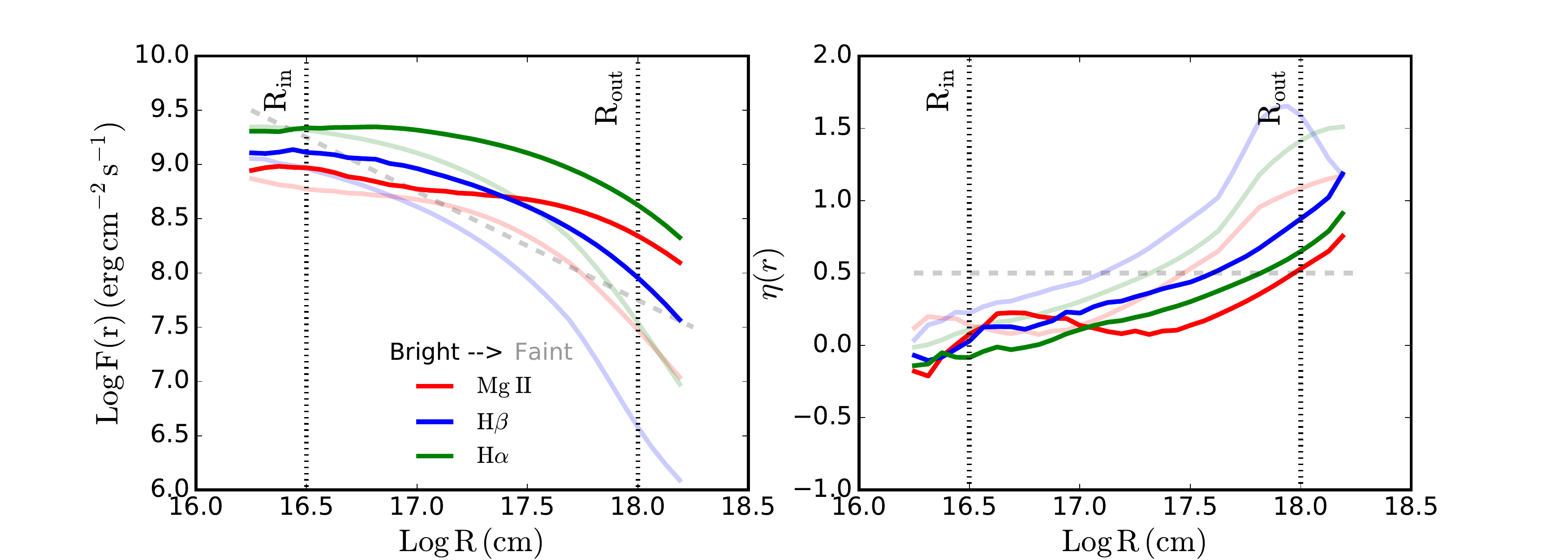}
\caption{Left: the radial emissivity function. Emission-line radial surface fluxes based on the weighting function along the gas density axis. Right: the radial responsivity function. $\eta$ is calculated based on Eqn.\ (\ref{eq:eta}). The gray dashed line of $F(r) \propto r^{-1}$ (left) indicates $\eta = 0.5$ (right).} The radial distance is calculated from the continuum luminosity $L_{\rm 3000\AA}$ = [44, 45] \erg (transparent and solid). The vertical dashed lines mark inner and outer boundaries of the BLR ($R_{\rm in}$ $\simeq$ 12 lt-day, $R_{\rm out}$ $\simeq$ 385 lt-day). 
\label{fig:Fr}
\end{figure*}

We sum the grid emissivity in Fig. \ref{fig:LOC} along the density axis at each radius to obtain the radial distribution of surface emissivity for different lines, as shown in the left panel of Fig. \ref{fig:Fr}. We note that only the density range of 8 $\leq$ $\rm log\, n(H)$ (\cmt) $\leq$ 12 is considered, since below $n(H) = 10^8$ \cmt the clouds are inefficient in producing emission lines and above $n(H) =10^{12}$ \cmt the clouds mostly produce thermalized continuum emission rather than emission lines \citep{Korista00}. In addition, we only sum the contributions from ionized clouds with $\rm 6 \le log\,U_{H}c \le 11.25$ \citep[see details in][]{Korista00,Korista04a,Korista19}.

\begin{figure*}
\centering
\includegraphics[width=18.cm]{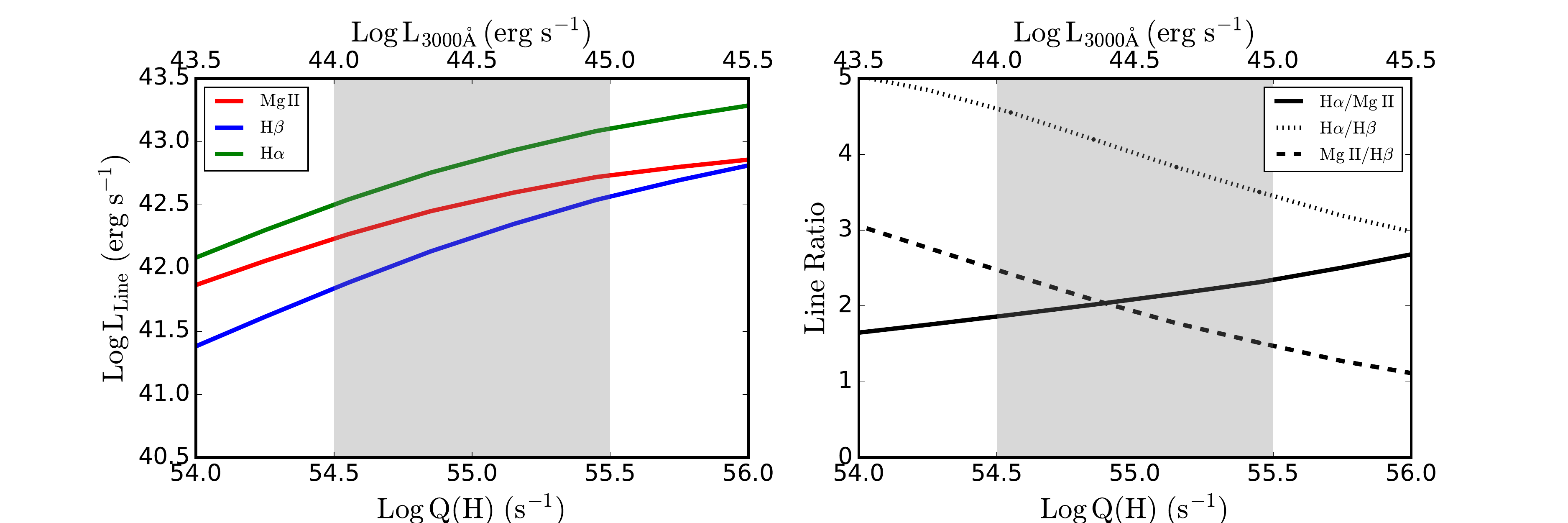}
\caption{Left: responses of line luminosity to continuum changes in the fiducial LOC model. The line lumimoisty is computed assuming a global covering fraction of $C_{f} = 50\%$, i.e., half of the continuum emission is covered by the clouds.} Right: line ratios as a function of continuum luminosity in the fiducial LOC model. The gray shaded regions enclose a factor of ten change in continuum luminosity to mimic a CLQ (or extreme variability quasar). 
\label{fig:model_LOC}
\end{figure*}

In the left panel of Fig. \ref{fig:Fr}, we also consider the case where the continuum luminosity at 3000 \AA\ of the quasar drops by a factor of 10, i.e., $L_{\rm 3000\AA}$ decreases from $10^{45}$ to $10^{44}$ \erg\ (gray shaded area in Fig. \ref{fig:model_LOC}), corresponding to Q(H) $\simeq$ [$10^{55.5}$, $10^{54.5}$] ($\rm s^{-1}$). When the quasar continuum changes from the bright state (solid lines) to the faint state (lighter lines) by one dex, the radial emissivity function simply shifts to the left by 0.5 dex (since $\Phi(H) \propto L/r^2$) assuming no dynamical structure changes of the BLR, as well as its inner and outer boundaries. \mgii\ emissivity changes much less than Balmer lines in our assumed BLR, and this difference in the level of line emissivity changes from the bright state to the faint state increases with increasing radius.

Following previous works \citep[e.g.,][]{Goad93,Korista04a}, we define a responsivity parameter 
\begin{equation}\label{eq:eta}
\rm \eta = \frac{dlogF(r)}{dlog\Phi(H)} \propto -0.5\frac{dlogF(r)}{dlog{r}}, \ since\ \Phi(H) \propto r^{-2},
\end{equation}
to describe the efficiency of converting the change in the ionizing continuum flux to the change in the responding line flux. Integrating over the full line-emitting region, we have the relation $L_{\rm line} \propto L_{\rm con}^{\eta}$. Furthermore, if we assume $F(r) \propto r^{\gamma}$ with $\gamma = -1$, then $\eta = -\gamma/2 = 0.5$, which is a rough approximation for several UV/optical emission lines \citep[e.g.,][also see the dashed gray lines in Fig. \ref{fig:Fr}]{Goad12}.

The right panel of Fig. \ref{fig:Fr} shows that the responsivities $\eta$ of Balmer lines and \mgii\ are correlated with the emitting radius and anti-correlated with the incident luminosity $L(\rm t)$ at different times. Since the clouds with $\eta(r, L(\rm t))> 1$ and $<0$ will not respond appropriately to the variations in the ionizing continuum \citep{Korista04a}, we generally consider the $0<\eta<1$ region. Since $\eta$ starts to exceed 0 at $R = 10^{16.5}$ cm for both the bright and faint states, we adopt $R_{\rm in} = 10^{16.5}$ cm (12 lt-day) for our LOC model. As long as the inner boundary is small, it has minor impact on our results because the clouds there have very high density ($> 10^{12}$ $\rm cm^{-3}$) and low responsivity $\eta$ with cloud emission dominated by continuum not emission lines.
However, the outer BLR boundary is important in our LOC model, which will determine the ``breathing" of the broad emission lines (see \S \ref{sec:breathing}). Note that although $\eta$ will be larger than 1 at $R_{\rm out} = 10^{18}$ cm in the faint state, most of these clouds are emitting inefficiently and will have little impact on our results.

To calculate the line luminosity, we still need to specify the distribution functions of clouds, i.e., $f(r)$ and $g(n)$. Traditionally LOC models have used empirical parameterizations for these distribution functions aiming at reproducing the observed emission line properties. For example, according to \cite{Baldwin95}, $f(r)\propto r^{\Gamma}$ and $g(n) \propto n^{\beta}$ ($\Gamma= -1$ and $\beta = -1$) are simple and reasonable assumptions for BLR clouds.  This parameterization of $f(r)$ and $g(n)$ results in equal weighting for each grid point in the density--flux plane in log-scale. For the best known NGC 5548, \cite{Korista00} suggested that $-1.4< \Gamma < -1$ is the optimised range to recover the observed time-averaged UV spectrum in 1993, and hence it was fixed to $-1.2$ in \cite{Korista19}. In this work we fix $\Gamma = -1.1$ to match the \mgii\ luminosity observed in a rare \mgii\ Changing-Look Quasar (CLQ, see the details in \S \ref{sec:cl}). Hence Eqn. (\ref{eq:Lline}) becomes
\begin{equation}\label{eq:Lline1}
L_{\rm line} \propto \int d(\log n) \int^{R_{\rm out}}_{R_{\rm in}} r^{1.9} F(r) d(\log r)
\end{equation}

Given the luminosity range of the quasar, we obtain the LOC predicted $L_{\rm con}$ -- $L_{\rm line}$ relation in Fig. \ref{fig:model_LOC}. In the left panel, \mgii\ varies at a slower rate with continuum than the Balmer lines, particularly in the gray region that encloses the typical luminosity range of a CLQ \citep[e.g.,][]{Macleod16,Yang18}. In this shaded region, when the continuum luminosity drops by 1 dex, the \mgii\ luminosity is only reduced by 0.45 dex, which is less than the luminosity reduction in Balmer lines (e.g., 0.6 dex for \ha\ and 0.7 dex for \hb). In the right panel, we show different line ratios as a function of continuum luminosity. With decreasing central luminosity, the ratios of \ha/\hb\ and \mgii/\hb\ increase, which is consistent with the theoretical results in \citet{Baldwin95}, and observations of quasar composite spectrum \citep{VandenBerk01} and CLQs \citep{Macleod16,Macleod19,Yang18}. 

\subsection{Reproducing the ``breathing" of broad lines}\label{sec:breathing}


Given the long dynamical timescale (more than a few decades) in the dust sublimation region, the physical outer boundary of the BLR is determined by the average luminosity state at least decades ago, and could be largely unrelated to the current luminosity state. Here we explore the possibility that $R_{\rm out}$ deviates from that estimated based on the current luminosity and the consequences on the breathing behaviors of different broad lines.

\begin{figure*}
\centering
\includegraphics[width=18cm]{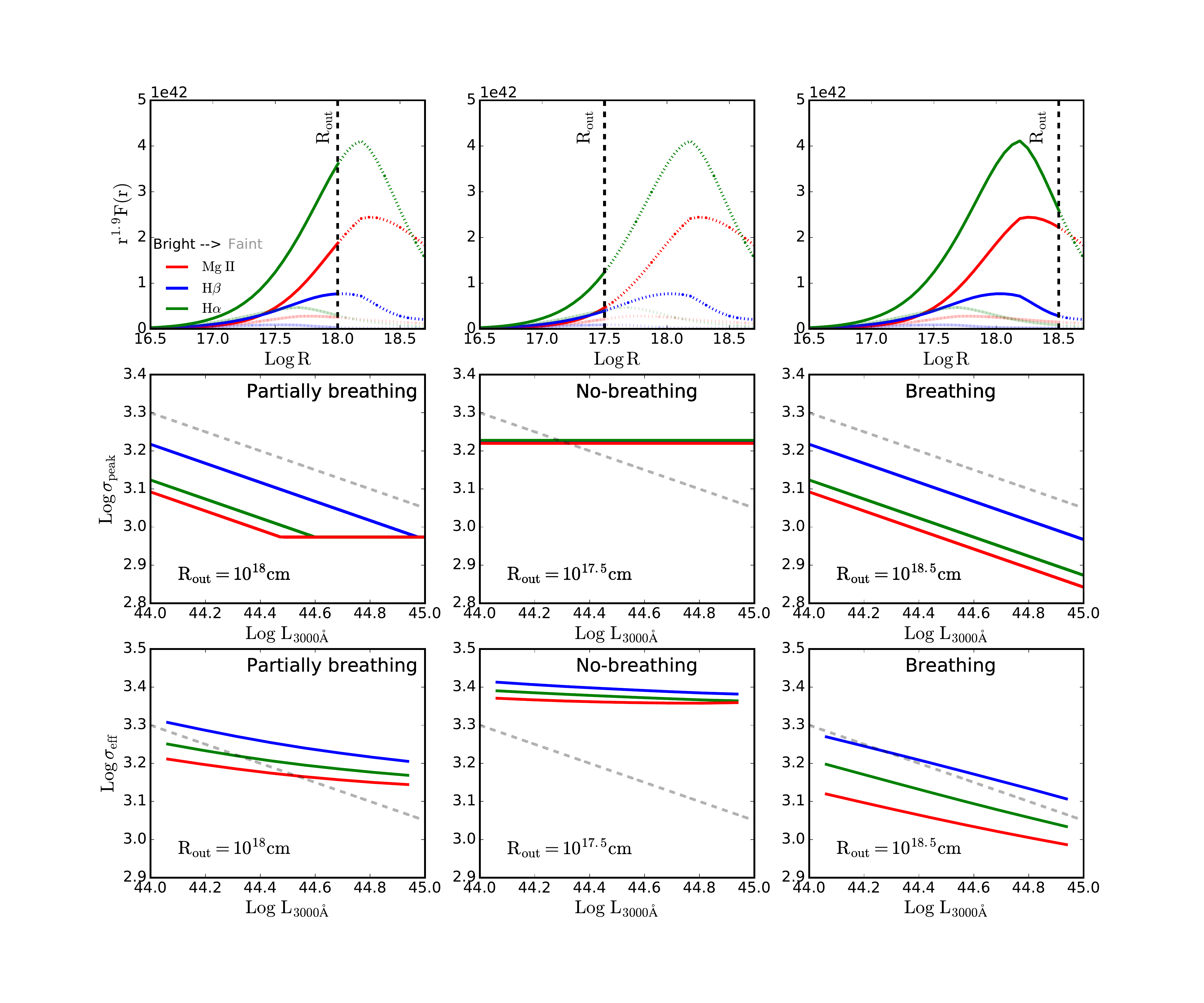}
\caption{The radial emissivity functions for different lines (upper panels) and the corresponding ($\sigma_{\rm peak}$-based \& $\sigma_{\rm eff}$-based) breathing modes (middle/lower panels). In the upper panels, the solid-dotted line corresponds to the bright state, and the lighter line corresponds to the faint state  with a factor of ten (1 dex) drop in flux. For the faint state, the line emissivity peaks move to the lower left accordingly. The dashed vertical line marks the outer radius of the BLR, and the dotted portions of the emissivity profile are not used in the calculation of line luminosity. Each column corresponds to a different outer radii $R_{\rm out}$ to demonstrate the resulting luminosity-line width ($L-\sigma$) relation. Perfect breathing corresponds to $L\propto R^2\propto \sigma^{-4}$(dashed gray lines).
}
\label{fig:emissivity}
\end{figure*}

If we consider the previous $L_{\rm 3000\AA}$ that set the outer BLR boundary is 10 times smaller or larger than the current average luminosity, we obtain $R_{\rm out} = 10^{17.5}$ or $10^{18.5}$ cm assuming $\rm log \Phi(H) = 17.9 \ \rm cm^{-2}\,s^{-1}$ and that the BLR outer boundary has not yet had enough time to dynamically adjust itself due to luminosity state changes. Thus with three values of $R_{\rm out} = 10^{17.5}, 10^{18} $ and $10^{18.5}$, we produce three cases that have different behaviors in the relation between line width and luminosity (e.g., 2rd \& 3rd rows of Fig. \ref{fig:emissivity}): Column (1) \hb\ shows ``breathing'' but \mgii\ and \ha\ only cover half of the luminosity range (partially breathing); Column (2) none of the lines show ``breathing'' (no-breathing); and Column (3) all three lines show ``breathing'' (see details below). 

The resulting radial distributions of line emission are presented in the upper panels of Fig. \ref{fig:emissivity}. Similarly we consider a continuum change of 1 dex from the bright state (solid lines) to the faint state (lighter lines) to mimic a CLQ. These radial emissivity profiles will not only shift to the left, but also move down due to $\Phi(H) \propto L/r^2$. The \mgii\ emissivity peak is always located at larger radii than those for the Balmer lines Moreover, the \mgii\ \& \ha\ emissivity profiles are more narrowly distributed than that of \hb. In Column (1), the maximum emissivity for all three lines is located very close to the outer boundary of the BLR in the bright state. In the faint state, the maximum emissivity of \hb\ is shifted to inside the outer boundary, while the maximum emissivities of \mgii\ and \ha\ are still near the outer boundary.

Assuming the BLR is virialized\footnote{Note that we do not have enough information to reconstruct the detailed kinematic structure of the BLR. It is possible that some portion of the BLR gas is not in a virialized component, which could contribute to the wings of the broad lines \citep[e.g.,][]{Ho12,Popovic19}.}, we compute the average virial velocity and the corresponding observed broad line width (line dispersion both for $\sigma_{\rm peak}$ and $\sigma_{\rm eff}$) for each line and display the results in the lower panels of Fig.\ \ref{fig:emissivity}.  The $\sigma_{\rm peak}$ is calculated based on the peak-emissivity radius $R_{\rm peak}$. This simplification is adopted to demonstrate the concept of different ``breathing" modes since the line emission is nearly dominated by the emissivity peak \citep{Baldwin95}. We also consider a more realistic line width estimation, the effective line width $\sigma_{\rm eff}$ weighted by the radial line emissivity:
\begin{equation}\label{eq:effsig2}
\sigma_{\rm eff} =  \frac{\int \sigma r^{1.9}F(r)dlog(r)}{\int r^{1.9}F(r)dlog(r)},
\end{equation}
where $\sigma = \sqrt\frac{M_{\rm BH}G}{Rf}$,  $M_{\rm BH} = 10^{8.5} M_{\odot}$ and $f = 4.47$.

As shown in Fig. \ref{fig:Fr}, the $\sigma_{\rm peak}$-based cases clearly present distinctions between breathing and no-breathing regions, and hence we use it to define three breathing categories. However, for more realistic situations, all $\sigma_{\rm eff}$-based cases show identical anti-correlations, but only with different slopes more skewing toward breathing or no-breathing scenarios. Both estimations demonstrate that \mgii\ is always the least breathing among the three lines for $R_{\rm out} = 10^{18}$ cm. Comparing to the typical observed line width ($\sigma$ $\sim$ 1700 \kms) for \hb\ in SDSS quasars with $M_{\rm BH} = 10^{8.5}M_{\odot}$, $\sigma_{\rm eff}$ ($\sim$ 1600 \kms) is preferred over the use of $\sigma_{\rm peak}$ ($\sim$ 1000 \kms). In addition, since Column (1) is the most consistent with observed properties for the Balmer lines and \mgii\ \citep[e.g.,][]{Park12,Shen13,Yang19}, we will use the $\sigma_{\rm eff}$-based partially breathing model as our fiducial model to produce further relations and CL sequence in \S\ref{sec:ewsig} \& \S\ref{sec:cl}. This model is the same fiducial LOC model for all other predictions.

\subsection{The relation between broad-line width and equivalent width}\label{sec:ewsig}

\begin{figure*}
\centering
\includegraphics[width=18.cm]{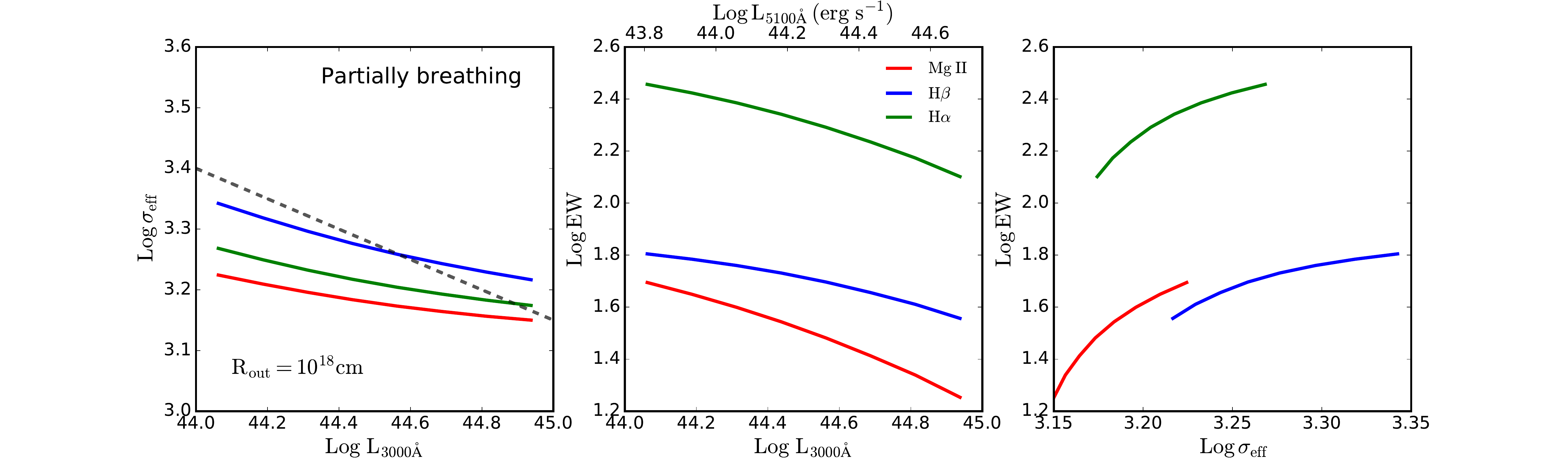}
\caption{The relations among $L_{\rm 3000\AA}$, EW and $\sigma_{\rm eff}$. The calculation is based on our fiducial partially breathing mode assuming $R_{\rm out}$ = $10^{18}$ cm. The dashed gray line indicates the perfect breathing model in the left. The EW of \hb\ and \ha\ is computed with the continuum flux at 5100 \AA.  
}
\label{fig:FWHM-EW}
\end{figure*}

The intrinsic Baldwin Effect \citep[BE,][]{Baldwin77} states that the line EW decreases with increasing continuum luminosity for a given quasar. From Eqn.\ (\ref{eq:eta}), 
\begin{equation}
    \rm \eta = \frac{dlogF(r)}{dlog\Phi(H)} = \frac{dlogEW}{dlog\Phi(H)}+1,
\end{equation}
and therefore $L_{\rm con} \propto EW^{\eta -1}$, which means the slope of the intrinsic BE is governed by the responsivity $\eta$ that varies with the continuum luminosity and the formation radius. For instance, the responsivity $\eta$ for \hb\ in partially breathing BLRs for the bright (faint) state is $\sim$ 0.5 (0.7) at the most effective formation radius $R_{\rm eff} = 10^{17.52}\ (10^{17.25})$ cm (see Figs. \ref{fig:model_LOC} \& \ref{fig:Reff}). Thus the BE slope varies between $-0.5$ and $-0.3$ from the bright to the faint states, which is consistent with the predication ($\sim -0.4$) in Fig. \ref{fig:FWHM-EW}. For the well-studied AGN NGC 5548, the intrinsic BE slope $\sim -0.6$ predicted by LOC models has been confirmed by observations \citep[e.g.,][]{Gilbert03,Rakic17}, verifying the reliability of the LOC model. In addition, the average EW values of different emission lines (i.e., $\rm logEW_{\rm MgII} = 1.5$ and $\rm logEW_{\rm \hb} = 1.8$ ) in Fig.\ \ref{fig:FWHM-EW} are also consistent with observed values for SDSS quasars around $L_{\rm 3000\AA} = 10^{44.5}$ \erg\ \citep{Shen11}. Note that the EWs for \hb\ and \ha\ are computed using $L_{\rm 5100\AA}$, scaled from $L_{\rm 3000\AA}$ by a factor of 1.8 based on the average quasar SED \citep{Richards06}.    

On the other hand, the model also predicts a correlation between the line width and the EW. This correlation is driven by the combined effect of breathing and the BE: both line width and EW are anti-correlated with continuum luminosity. However, if \mgii\ is partially breathing in the luminosity range probed in Fig. \ref{fig:FWHM-EW}, it predicts a steeper slope of $\sim 4$ for \mgii\ in logarithm space than the observed slope ($\sim 1$) in population studies \citep{Dong09,Shen11}. It is likely that the relations revealed here for a single quasar (i.e., the relation is intrinsic) contribute to the global relation observed for populations of quasars with different BH masses and luminosities.  

\begin{figure}
\centering
\includegraphics[width=10.cm]{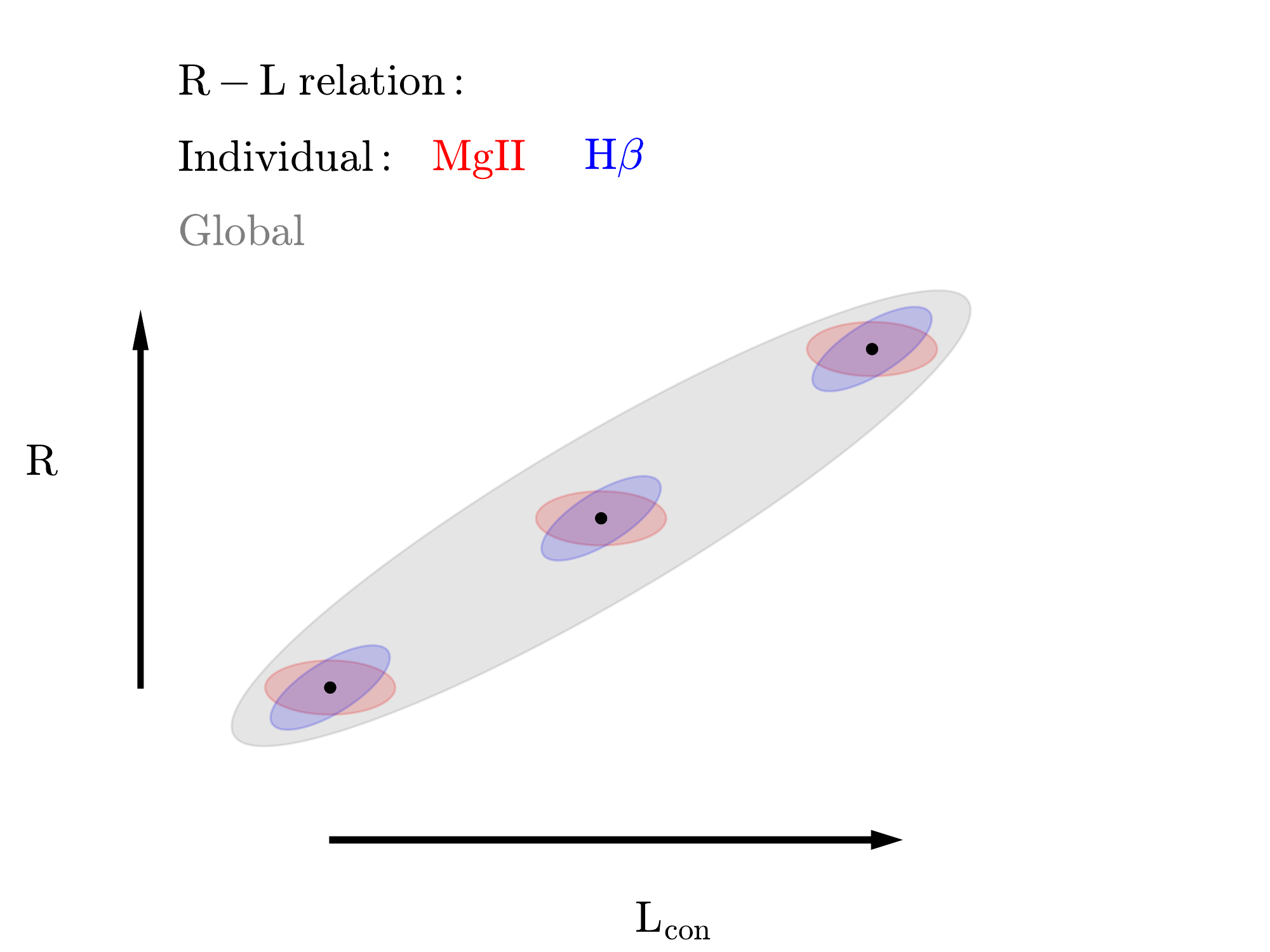}
\caption{A cartoon for the $R-L$ relation. The gray ellipse represents a global $R-L$ relation for the population of quasars with different BH masses and luminosities. The small blue and red ellipses represent the intrinsic $R-L$ relation for a single quasar with variable luminosity, for \hb\ and \mgii, respectively. The ``breathing" mode for \hb\ in individual quasars is the result of the intrinsic $R-L$ relation. For \mgii\ there is no intrinsic $R-L$ relation, which results in no ``breathing'' mode for \mgii.}  
\label{fig:R-L}
\end{figure}

\begin{figure*}
\centering
\includegraphics[width=18.cm]{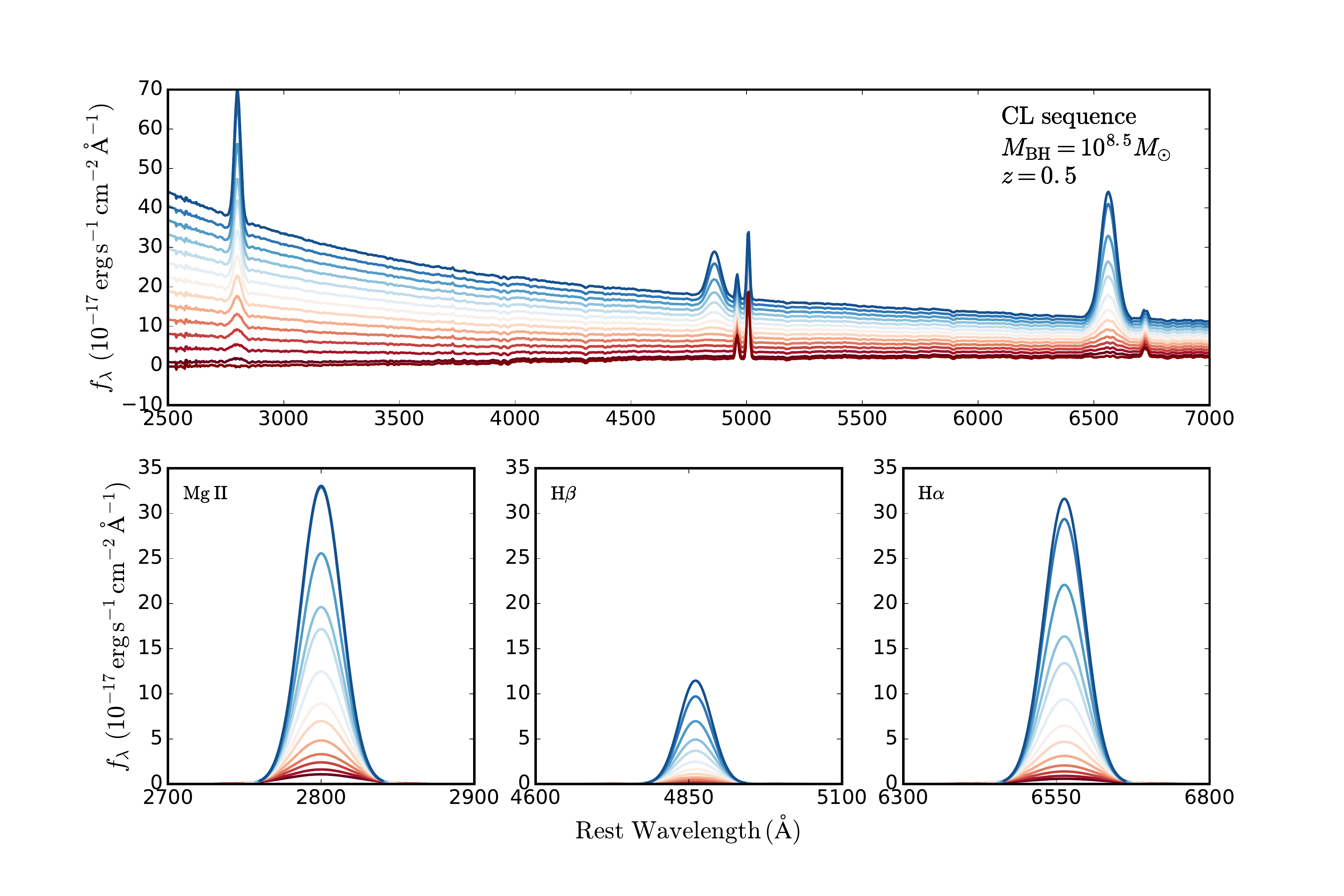}
\caption{A CL sequence for quasars. Lower panels: theoretical line profiles of \mgii, \hb\ and \ha\ predicted by the fiducial LOC model assuming a quasar with $M_{\rm bh} = 10^{8.5}\,M_{\odot}$ at $z = 0.5$. The line flux and width are governed by the $L_{\rm line}$ -- $L_{\rm con}$ and $L_{\rm con}$ -- $\sigma$ relations in Fig. \ref{fig:model_LOC} and the partially breathing case in Fig. \ref{fig:emissivity}. All three lines are simulated by single Gaussians. Upper panel: $L_{\rm 3000\AA}$ is reduced from $10^{45}$ to $10^{43}$ \erg for the full spectrum in steps of 0.15 dex. Each spectrum consists of a power-law continuum, a host galaxy template, and Gaussians of broad and narrow lines except for the faintest epoch, which only contains host emission and narrow lines. The second faintest epoch has a \mgii\ luminosity tuned to $\sim 10^{41.5}$ \erg\ with $\Gamma = -1.1$, which is consistent with the observed line luminosity in the \mgii\ CLQ reported in \cite{Guo19}. In each panel, the color gradient represents luminosity changes from the brightest epoch (blue) to the faintest epoch (red).
}
\label{fig:cl_seq}
\end{figure*}

\begin{figure}
\centering
\includegraphics[width=9.cm]{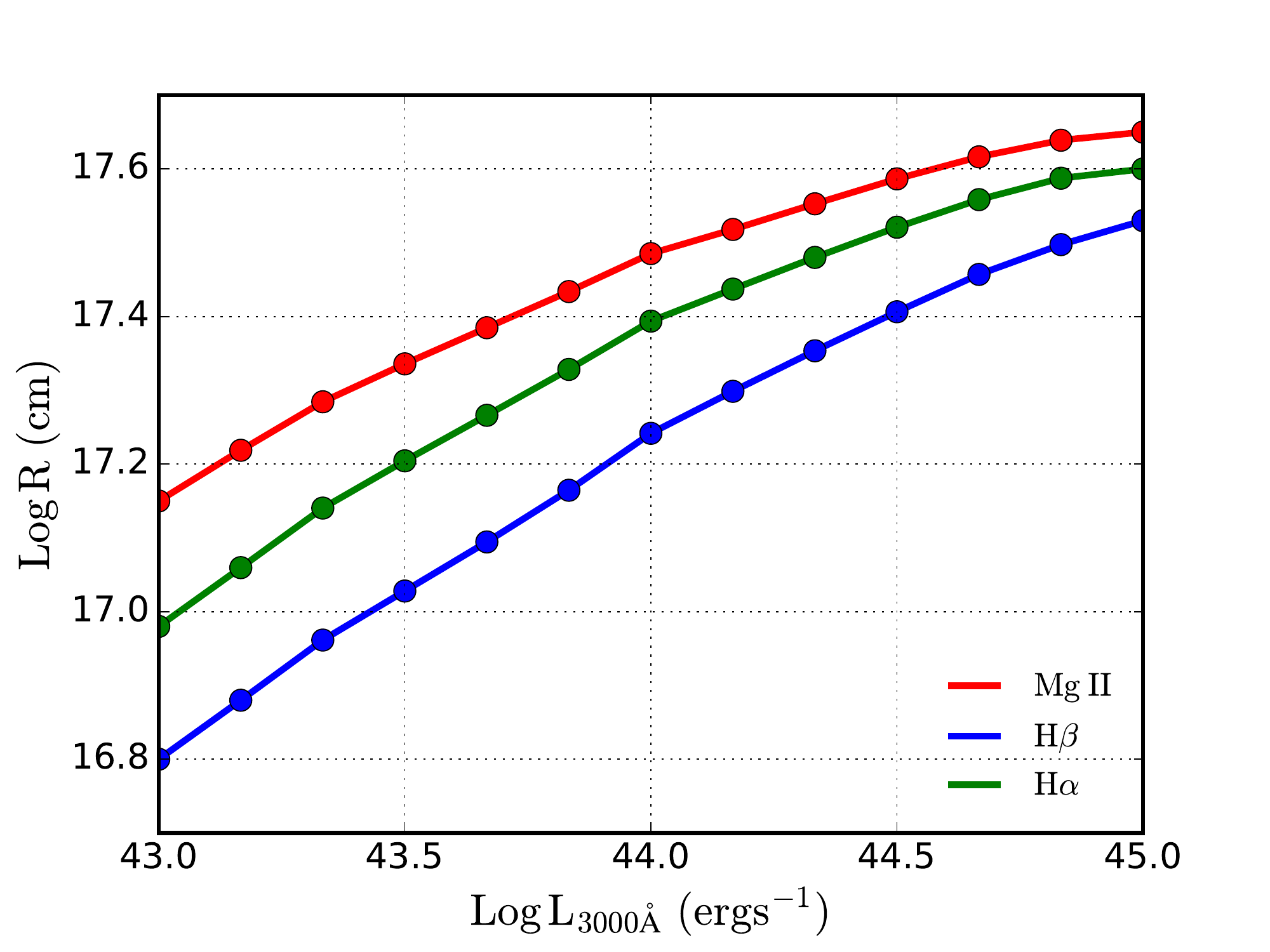}
\caption{Average line-emitting radius for each epoch in the CL sequence shown in Fig.\ \ref{fig:cl_seq}, estimated using the emissivity-weighted line width $\sigma_{\rm eff}$. The average radius of \mgii\ decreases more slowly than the Balmer lines.     
} 
\label{fig:Reff}
\end{figure}

\section{Implications}\label{sec:imp}


The appeal and caveats of the LOC model have been discussed extensively in earlier work \citep[e.g.,][]{Korista00,Korista04b}. While the LOC model is quite successful in reproducing the bulk of the observed properties of the broad line emission, we fully acknowledge the empirical nature of this approach. For example, the modeling of the covering factors (weights) as functions of radius and cloud density is somewhat ad hoc. Nevertheless, we found that this empirical photoionization model can explain most of the observed variability properties of the \mgii\ line and their differences from those for the Balmer lines. 

Assuming that our fiducial LOC models are the correct prescription for broad line emission in quasars, we discuss the implications of the predicted broad line variability, with an emphasis on the \mgii\ line.  

\subsection{\mgii\ reverberation mapping}

As discussed in \S\ref{sec:intro}, \mgii\ is an important broad line of RM interest at intermediate redshifts with optical spectroscopy. Confirming earlier photonionization calculations \citep[e.g.,][]{Korista00,Korista04a}, we found that the variability properties of broad \mgii\ can be reasonably well explained by the LOC model. First, the broad \mgii\ responses to the continuum variability at a lower level than the broad Balmer lines, which means that the \mgii\ lags will be more difficult to detect in general. The fact that the \mgii\ gas on average is located at slightly larger distance than the Balmer line gas also adds to the difficulty of detecting \mgii\ lags, since longer monitoring duration is required to capture the lag. Moreover, there is a general mismatch between the formation radius of the lines and the characteristic timescale of the driving continuum which means that in general the measured delays, and thus inferred ``sizes'', are underestimated 
 \citep{Perez92,Goad14}.

Perhaps a more striking feature of \mgii\ variability is the general lack of breathing when luminosity changes. We have argued that this lack of breathing could be explained by the possibility that the \mgii\ gas with maximum emission efficiency is near the outer physical boundary of the BLR, and hence the flux-weighted radius of the \mgii\ emitting clouds does not vary much with luminosity\footnote{Under different model parameters, \mgii\ can also exhibit breathing (Fig.\ \ref{fig:emissivity}), as observed in rare cases \citep{Dexter19a}.}. Alternatively, if the inner and outer boundaries of the BLR are fixed and $\eta$(r, L(t)) is constant over radius, which means the responsivity $\eta$ is the same for the line core and wings, the line will not display breathing as the entire profile scales up and down with continuum variation \citep[e.g.,][]{Korista04b}. A third possibility is that if the outer part of the BLR is dominated by turbulent motion (or non-virial motion), the line width of \mgii\ will also be less sensitive to continuum variations \citep[e.g.,][]{Goad12}. The lack of breathing for \mgii\ suggests that there is no intrinsic $R-L$ relation for the \mgii\ BLR. However, a global $R-L$ relation may still exist for quasars over a broad BH mass and luminosity range, if the outer radius $R_{\rm out}$ scales with BH mass. Fig. \ref{fig:R-L} demonstrates the possible existence of a global $R-L$ relation and the absence of an intrinsic $R-L$ relation for \mgii. More RM results on \mgii\ will be important to test the existence, or lack thereof, of a global $R-L$ relation \citep[e.g.,][]{Shen15,Shen16,Czerny19}. 

One assumption in our photoionization modeling is that the physical BLR structure does not change over the period of the continuum variability. It is possible that $R_{\rm out}$ will slowly change on dynamical timescale at this radius ($> 10^2$ yr for our default parameters of $M_{\rm BH}=10^{8.5}\,M_\odot$ and $R_{\rm out}\sim 0.3$ pc), and may eventually settle down at a different value in a new average luminosity state over the much longer lifetime of the quasar. This possibility could also be responsible for a global $R-L$ relation for \mgii.


\subsection{A changing-look sequence}\label{sec:cl}

Our LOC models also have implications for the behaviors of broad-line responses to continuum changes in the population of optically-identified CLQs, or more generally, extreme variability (or hypervariable) quasars \citep[e.g.,][]{Rumbaugh18}.   

Most of the CLQs reported so far are spectroscopically defined by dramatic changes in the broad Balmer line flux between the bright and the dim states \citep[e.g.,][]{LaMassa15,Runnoe16,Macleod16,Macleod19,Sheng17,Yang18}. In most, if not all, of these cases, broad \mgii\ remains visible even in the dim state \citep[e.g.,][]{Macleod19,Yang19}. 

With our LOC models, we can qualitatively explain the persistence of broad \mgii\ in a CL event. Fig. \ref{fig:cl_seq} presents a time sequence of synthetic spectra with the continuum luminosity decreasing from $10^{45}$ to $10^{43}\, {\rm erg\,s^{-1}}$, for the fiducial model ($R_{\rm out}= 10^{18}$ cm) that can reproduce the observed \mgii\ variability properties (see \S\ref{sec:weakmgii} and \S\ref{sec:breathing}). Each spectrum consists of a quasar power-law continuum, a host galaxy component\footnote{The host galaxy template is taken from the actual spectrum of a recently-discovered \mgii\ changing-look object \citep{Guo19}}, and broad/narrow emission lines (e.g., \ha, \hb, \mgii, \OIII\ and \SII) described by single-Gaussian functions. In the faintest state we only include the host galaxy and narrow emission lines. We adopt a quasar continuum power-law slope $\alpha=-1.56$ ($f_{\rm \lambda}=\lambda^{\alpha}$) \citep[e.g.,][]{VandenBerk01}. We start at the brightest epoch with $L_{\rm 3000\AA}=10^{45}$ \erg, and reduce the continuum luminosity in steps of 0.15 dex. The strengths of the broad emission lines are computed from our fiducial LOC model at each continuum luminosity. Note that the index $\Gamma$ for the radial distribution of cloud coverage determines the line luminosity and the BH mass governs the line dispersion (since the effective radius is determined from photonization). For our fiducial LOC model, the $M_{\rm BH}$ is fixed to $10^{8.5}M_{\odot}$, which yields $\sigma_{\rm eff}$ = $10^{3.4}$ \kms at continuum luminosity of $10^{43}$ \erg. Meanwhile, \mgii\ luminosity is required to be $\sim$ $10^{41.5}$ \erg\ at this continuum luminosity to match the only reported \mgii\ CLQ in \citet{Guo19}, which led to the choice of $\Gamma$ = $-1.1$ in \S\ref{fig:LOC} for our fiducial LOC model.

From the sequence in Fig. \ref{fig:cl_seq}, it is obvious that when the broad Balmer lines almost disappear (e.g., become undetectable), the broad \mgii\ emission remains visible. When the continuum luminosity continues to drop, broad \mgii\ will eventually become too weak to be detectable. Indeed we have found an example of a \mgii\ CL object from a systematic search with repeated SDSS spectra \citep{Guo19}. In that case, the \mgii\ equivalent width dramatically changed from $\sim$100 \AA\ to being consistent with zero with little continuum change due to the dominance of the host light at the dim state. In Fig. \ref{fig:Reff}, we calculate the $\sigma_{\rm eff}$-based average radius for each epoch in this CL sequence, which decreases when luminosity decreases. Most importantly, the average radius (line width) of \mgii\ decreases (increases) more slowly than those of the Balmer lines, verifying our fiducial model in Figs.\ \ref{fig:emissivity} \& \ref{fig:FWHM-EW}.

We note that there is currently some ambiguity in the observational definition of ``CLQs'' based on the appearances of the broad Balmer lines, i.e., the detection of broad Balmer lines is S/N dependent. Our LOC calculations demonstrate that the variability of the broad emission lines in a CLQ can be fully explained by photonization responses to the dramatic continuum changes. There is nothing special about the properties of the broad emission lines in CLQs compared to normal quasars, except for their extreme continuum variability. For these reasons, ``extreme variability quasars'', or ``hypervariable quasars'', is a more appropriate category term for these objects in our opinion.    

Given the sequence of broad-line spectra shown in Fig. \ref{fig:cl_seq} following the fading of continuum emission, it is possible to catch the quasar in a state where there is detectable broad \mgii\ but no detectable broad Balmer lines. \citet{Roig14} discovered that there is a rare population of broad \mgii\ emitters in spectroscopically confirmed massive galaxies from the SDSS. We postulate that these broad \mgii\ emitters may be the transition quasar population where the quasar continuum and broad Balmer line flux had recently dropped by a large factor but the broad \mgii\ flux is still detectable on top of the stellar continuum. Using the sample of broad \mgii\ emitters from \citet{Roig14}, we confirmed that the EWs of the broad \mgii\ follow the extrapolated Baldwin effect in our LOC model to lower luminosity range\footnote{We found that the 3000\AA\ luminosity for these broad \mgii\ emitters were underestimated by an erroneous factor of 100 in the original Roig et al. paper (B. Roig, private communications).} of $L_{3000}\sim 10^{43} {\rm erg\,s^{-1}}$. Thus the LOC model provides a natural explanation for these rare broad \mgii\ emitters in otherwise normal galaxy spectra.

\subsection{Comparison with a well-studied case: NGC 5548}

NGC 5548 is the ideal case to compare with our LOC model predictions given its extensive optical/UV RM data in the past several decades. We collected the published RM results from the 13-year AGN Watch project \citep{Peterson04} and the 6-month Space Telescope and Optical Reverberation Mapping project (AGN STORM) project \citep{Pei17} to investigate the relations between continuum and emission line fluxes (i.e., \hb). Unfortunately for NGC 5548 the data on \mgii\ are much less than the data on \hb\ and therefore we do not consider \mgii\ here.

\begin{figure*}
\centering
\includegraphics[width=18.cm]{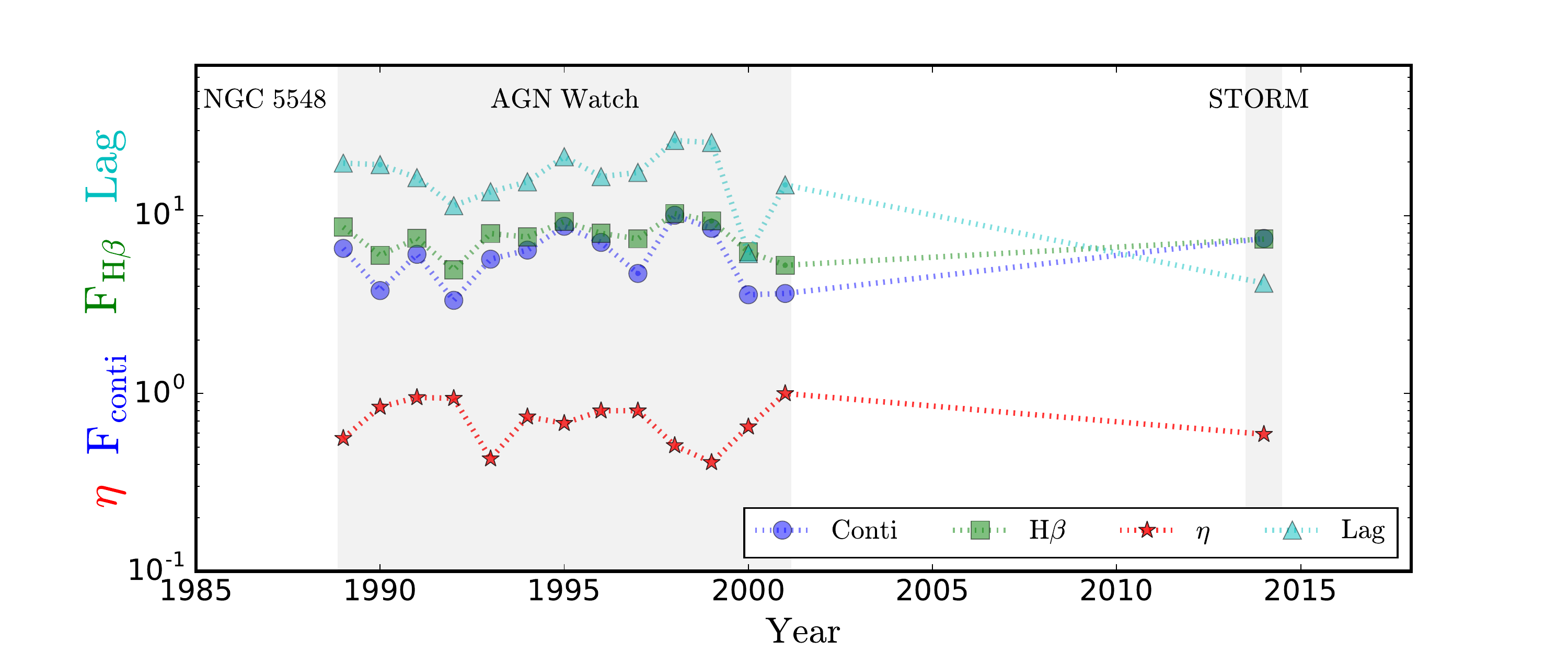}
\caption{Time evolution of the continuum flux at 5100 \AA\ in units of $10^{-15}$ \ergcmsA, broad \hb\ flux in units of  $10^{-13}$ \ergcms, \hb\ time lag in units of days, and responsivity ($\eta$). The data from 1988-2001 and 2014 are taken from the AGN Watch \citep{Peterson04} and STORM \citep{Pei17} projects, respectively. The host contribution and narrow-line \hb\ flux were removed when calculating the continuum and broad-line fluxes.
} 
\label{fig:RM}
\end{figure*}

As shown in Fig.~\ref{fig:RM}, the continuum flux (host-corrected) at 5100 \AA\ and line flux are well correlated over several decades \citep[also see][]{Goad14}. The time lags (or emitting size) between the continuum and \hb\ display the expected breathing during the AGN Watch program \citep[e.g,][]{Gilbert03,Goad04,Cackett06}. At the end of AGN Watch around 2000, NGC 5548 was in a historic low-state for some considerable time before returning to the average historic luminosity in the more recent AGN STORM campaign. If the outer BLR boundary was set during the low-state around 2000, we would expect the 2014 AGN STORM RM results to have weaker \hb\ strength and a reduced lag (since the BLR had been physically truncated), consistent with the restuls around 2000 (from AGN Watch) and in 2014 (STORM). Therefore it is plausible that there is a physical outer boundary of the BLR set by the prior average luminosity state, as postulated in our model.


\section{Conclusions}\label{sec:con}

In this work we have studied the variability of broad \mgii, \ha, and \hb\ in the framework of photoionization of BLR clouds by the ionizing continuum from the accretion disk in quasars. We adopt the popular but empirical LOC model with typical parameters used in the literature to qualitatively reproduce the observed variability properties of the three low-ionization broad lines. 

Our main findings are:

\begin{enumerate}
\item the LOC model confirms that the emissivity-weighted radius decreases in the order of \mgii, \ha\ and \hb\  (Figs. \ref{fig:LOC} \& \ref{fig:emissivity}), which is consistent with limited RM results where more than one lines have detected lags \citep{Clavel91,Peterson99,Shen16,Grier17} and previous photoionization predictions from \cite{Goad93,Korista00,Korista04a}. It also predicts that the \hb-emitting gas is more broadly distributed radially than \mgii\ and \ha\ (Fig. \ref{fig:emissivity}). 

\item the observed weaker variability and slower response of \mgii\ compared to the Balmer lines are recovered over a broad range of quasar continuum variations (Fig. \ref{fig:model_LOC}), which is again consistent with previous studies \citep[e.g.,][]{Goad93,Korista00}. These results come naturally from photoionization calculations that capture various excitation mechanisms and radiative transfer effects. Variability dilution due to the larger average distance of \mgii\ gas likely also contributes to this difference between \mgii\ and Balmer line variability. 

\item the general lack of ``breathing'' of the broad \mgii\ line can be explained by the possibility that the most efficient \mgii-emitting clouds are always near the outer physical boundary of the BLR. On the other hand, the Balmer line gas is inside this outer BLR boundary and the average line formation radius shifts as continuum luminosity changes to produce the ``breathing'' effect (Fig.~\ref{fig:emissivity}). Under certain circumstances when the \mgii\ gas is also mostly inside the outer BLR boundary, \mgii\ can also display the ``breathing'' behavior.    

\item based on these photoionization calculations, there is a natural sequence of the successive weakening of \hb, \ha, and \mgii, when the ionizing continuum decreases. The ``changing-look'' behavior in CLQs can be fully explained by the photoionization responses of the broad emission lines to the extreme variability of the continuum, adding to mounting evidence that most CLQs are caused by intrinsic accretion rate changes. Our results provide natural explanations for the persistence of broad \mgii\ line in CLQs, and broad \mgii\ emitters in otherwise normal galaxies. 
\end{enumerate}

The success of reproducing most of the observed \mgii\ variability properties, which only became available recently for statistical samples, with simple LOC models suggests that photonionization is the dominant process that determines the observed variability properties of the broad line emission. Future more RM results on \mgii\ will further test the LOC photoionization model, and to confirm the existence of a global $R-L$ relation for \mgii, a prerequisite to using the \mgii\ line as a single-epoch virial BH mass estimator for quasars.

\acknowledgments
We thank the referee for a thorough report and many suggestions that greatly improved this work. We also thank B. Roig for useful discussions on broad \mgii\ emitters, Luis Ho, Jon Trump and Gary Ferland for comments on the manuscript, and the CLOUDY team for valuable suggestions. Y.S. acknowledges support from an Alfred P. Sloan Research Fellowship and NSF grant AST-1715579. Z.H. is supported by NSFC-11903031 and USTC Research Funds of the Double First-Class Initiative. T.W. is supported by NSFC-11833007. M.K. is supported by Astronomical Union Foundation under grant No. U1831126 and Natural Science Foundation of Hebei Province No. A2019205100.




\bibliography{ref}

\end{CJK}

\end{document}